\documentstyle[prd,aps,multicol,graphicx]{revtex}

\newcommand{\be}{\begin{equation}} \newcommand{\ee}{\end{equation}}

\newcommand{\ba}{\begin{eqnarray}}
\newcommand{\ea}{\end{eqnarray}}

\newcommand{\bq}{\begin{equation}}
\newcommand{\eq}{\end{equation}}
\newcommand{\bqa}{\begin{eqnarray}}
\newcommand{\eqa}{\end{eqnarray}}
\newcommand{\ben}{\begin{enumerate}}
\newcommand{\een}{\end{enumerate}}
\newcommand{\bc}{\begin{center}}
\newcommand{\ec}{\end{center}}
\newcommand{\bqb}{\begin{eqnarray*}}
\newcommand{\eqb}{\end{eqnarray*}}

\begin{document}
\draft
\preprint{PM/04-7, HEPHY-PUB 787/04}

\title{\vspace{1cm}  Supersymmetry Tests from a Combined Analysis of
Chargino and Charged Higgs Boson Pair Production at a 1 TeV Linear Collider 
\footnote{Partially supported by EU contract HPRN-CT-2000-00149}}

\author{M. Beccaria$^{a,b}$, H. Eberl$^c$, 
F.M. Renard$^d$ and C. Verzegnassi$^{e, f}$ 
}

\address{
$^a$Dipartimento di Fisica, Universit\`a di
Lecce \\
Via Arnesano, 73100 Lecce, Italy.\\
$^b$INFN, Sezione di Lecce\\
$^c$Institut f\"ur Hochenergiephysik der \"Osterreichischen Akademie
der Wissenschaften,\\ A-1050 Vienna, Austria\\
$^d$ Physique
Math\'{e}matique et Th\'{e}orique, UMR 5825\\
Universit\'{e} Montpellier
II,  F-34095 Montpellier Cedex 5.\hspace{2.2cm}\\
$^e$
Dipartimento di Fisica Teorica, Universit\`a di Trieste, \\
Strada Costiera
 14, Miramare (Trieste) \\
$^f$ INFN, Sezione di Trieste\\
}

\maketitle

\begin{abstract}
We consider the production of chargino and charged Higgs boson
pairs at future linear colliders for c.m. energies in the one TeV
range. Working in the MSSM under the assumption of a "moderately"
light SUSY scenario, we compute the leading (double) and next-to
leading (linear) supersymmetric logarithmic terms of the 
so-called "Sudakov expansion" at one loop. 
We show that a combined analysis of the slopes of the chargino and of
the charged Higgs production cross sections would offer a simple possibility 
for determining $\tan\beta$ for large ($\gtrsim 10$) values and  an
allowed strip in the ($M_2,\mu$) plane. This could provide a strong
consistency test of the considered supersymmetric model.
\end{abstract}


\begin{multicols}{2}
\narrowtext

\section{Introduction}

One of the goals of a future electron-positron Linear
Collider \cite{LC} will be the performance of precision tests, hopefully of
models of new physics beyond the Standard Model, under the assumption
that a preliminary direct discovery of, at least, some of the new
particles requested by the candidate model has been achieved at
TEVATRON and/or LHC. These precision tests would then be particularly
relevant for models that involve a large number of parameters, some of
which might be different from the produced
particle masses and might turn
out to be determined with a relatively low accuracy. In alternative,
a number of competitor models might exist, showing a certain number of
properties (e.g. an identical set of particles), but differing in a
number of {\em fine} details that might remain invisible, or partially
hidden, in the process of direct production. In these cases, the need
of a precision test of the candidate model would become mandatory,
and should have the same relevance that the memorable precision tests
performed at LEP and SLC \cite{LEPSLC} had for a spectacular
confirmation of several predictions of the electroweak component of
the Standard Model.

If Supersymmetry were discovered at hadron colliders, the previous
discussion would immediately apply to the simplest existing
supersymmetric model, the MSSM. Being of perturbative nature, this
model would offer a large number of possibilities of performing
precision tests via the observation (and confirmation) of the virtual
effects that it predicts at the simplest higher perturbative order,
the so-called one loop level. Due to  the rather large number of
parameters that the model involves 
these tests would require the preparation and the use of
dedicated theoretical calculations and numerical programs, that are
actually already available in the literature.
As a general remark, it can be noticed that a general feature of these
programs, and of related precision tests as well, is that they are not
{\em simple}, and require dedicated efforts and attention.

The aim of this short paper is to show that, for a
{\em reasonable} light SUSY scenario  and for
c.m. energy values in the one TeV range, 
a rather simple test of the MSSM might be performed. 
This would imply the combined analyses of the production cross
sections of chargino and charged Higgs boson pairs at a number of
different energies in the one TeV region, in particular a 
determination of the slopes of
the cross sections.  The output of the effort would be, at a realistic optimal
experimental accuracy, the {\em determination} (a) of $\tan\beta$ with an
accuracy that would increase with increasing values of this parameter,
and (b) of an allowed strip in the plane of the two quantities
($M_2,\mu$), independently of the values of the remaining SUSY
parameters of the chosen scenario. The details of this
possibility will briefly be discussed in what follows.

\section{Determination of the logarithmic SUSY Sudakov terms}

Our analysis will be devoted to the two processes of 
production of charged Higgs and of chargino pairs in the
MSSM with real input parameters. For the first process, a complete calculation  
has been given at one-loop in two papers\cite{ChargedHollik,ChargedUs},
that is
supposed to be valid for arbitrary values of energy 
and parameters. To perform the calculations presented in 
  the second paper
\cite{ChargedUs}  the authors have also completed a numerical program (SESAMO).
The latter has been used to test the
 validity of a logarithmic Sudakov expansion in a scenario of {\em reasonably} light 
SUSY, i.e. one in which all the
 relevant SUSY masses are sufficiently smaller than the chosen c.m. energy. This 
has been fixed at about one TeV
 with a corresponding limit on masses of about 400 GeV, since from previous studies 
\cite{NewReality} one knows that in this
 configuration a one-loop logarithmic expansion does not require higher order 
corrections. From a comparison of the
 exact SESAMO results with the Sudakov logarithmic expansion it was concluded 
that the latter provides a
 satisfactory description of the process in the chosen scenario, with the only 
request of an additional constant (i.e.
 energy independent)
 term in the expansion, that will depend in general on the several 
SUSY parameters of the model in a way that it was tried to analyze in
 a qualitative stage. 
On the contrary, the SUSY loop contribution consists only of linear
logarithms stemming from Yukawa interactions. This term depends only on
the SUSY parameter $tan\beta$.
Therefore, a determination of 
the slope of the cross section,
 where the unknown SUSY constant term disappears (the other double and linear 
Sudakov logarithms are of pure SM
 origin and considered as known, together with the linear logarithms of RG 
origin for which the SUSY dependence is
 given by existing formulae), could provide a competitive~\cite{Djouadi,Ferrari}
determination of $\tan\beta$. Since a detailed description of this analysis has 
been given \cite{ChargedUs} we shall
 only write in this paper the relevant one-loop expressions for the cross section
of the process. Following the notations of \cite{ChargedUs} we 
shall thus write for the charged Higgs case ($\sqrt{q^2}$ is the c.m. energy):
\bq
\sigma^{\rm Born+1 loop} = \sigma^{\rm Born}(1+\Delta(q^2))
\eq
where in $\sigma^{{\rm Born+1 loop}}$ we are retaining only the genuine Sudakov 
one loop terms. The logarithmic expansion of $\Delta$ is 
given by the expression
\bqa
\Delta(q^2)&=& -~{\alpha\over2\pi s^2_W}
{1+2s^4_W\over1+4s^4_W}
\log^2{q^2\over M^2_W}\nonumber\\
&&-~{\alpha\over4\pi s^2_Wc^2_W}{1+2s^4_W+8s^6_W\over1+4s^4_W}
\log^2{q^2\over M^2_Z}\nonumber\\
&&-~{3\alpha\over4\pi s^2_WM^2_W}(m^2_t\cot^2\beta+m^2_b \tan^2\beta)
\log{q^2\over m^2_t}\nonumber\\
&&+~{\alpha\over3\pi s^2_Wc^2_W}
{11-16s^2_W+32s^4_W+72s^6_W\over1+4s^4_W}
\log{q^2\over M^2_Z}\nonumber
\label{Delta}
\eqa
where the third line contains the $\tan\beta$ dependent logarithms 
of Yukawa origin and we have neglected any {\em remainder} terms not 
growing with energy.
  
For the process of chargino pair production, a complete one-loop calculation is 
also available \cite{CharginoProduction}. In
this preliminary paper, we shall be interested in the study of the same scenario 
that was considered for the Higgs
pair production study previously discussed. In this spirit, we shall thus 
compute the two leading terms of the logarithmic
SUSY expansion and assume, supported by the mentioned study, that the only extra 
term to be added to the expansion is,
again, a (SUSY dependent) constant. We cannot, at this stage, prove this 
statement as we did for charged Higgs production; it appears
rather reasonable to us, and we shall devote a more complete following paper to 
a detailed investigation of its
validity. For the moment, we shall consider it as a working Ansatz, and show 
which  relevant consequences might be
derived under our assumption.

An essential difference between the processes of chargino and of charged Higgs 
pair production appears already at the
Born level. In the process $e^+ e^- \to \tilde \chi_i^+ \tilde \chi_j^-$,
the analytic expression of the scattering amplitude
contains already at lowest order the SUSY parameters $M_2$, $\mu$ and $\tan\beta$
appearing in the chargino mixing matrices $Z_{ij}^\pm$ (our notation is described 
in~\cite{NewReality})
\bqa
\lefteqn{A^{Born}_{ij} = -~{e^2/(4s^2_Wc^2_W s)}\{ } &&  \\
&& (\gamma^{\mu}P_{L})_{ee}.
(\gamma^{\mu}P_{L})_{\chi\chi}
[\delta_{ij}+(1-2s^2_W)Z^{+*}_{1i}Z^{+}_{1j}]\nonumber\\
&&+(\gamma^{\mu}P_{R})_{ee}.
(\gamma^{\mu}P_{R})_{\chi\chi}
[2s^2_W(\delta_{ij}-Z^{-}_{1i}Z^{-*}_{1j})]\nonumber\\
&&+(\gamma^{\mu}P_{L})_{ee}.
(\gamma^{\mu}P_{R})_{\chi\chi}
[\delta_{ij}+(1-2s^2_W)Z^{-}_{1i}Z^{-*}_{1j}]\nonumber\\
&&+(\gamma^{\mu}P_{R})_{ee}.
(\gamma^{\mu}P_{L})_{\chi\chi}
[2s^2_W(\delta_{ij}-Z^{+*}_{1i}Z^{+}_{1j})]\}
\nonumber\\
&&
-~{e^2/(2s^2_W)}[~{Z^{+*}_{1i}Z^{+}_{1j}/u
}~]~(\gamma^{\mu}P_{L})_{ee}.
(\gamma^{\mu}P_{L})_{\chi\chi} \nonumber
\eqa
where $s\equiv q^2$ and $u$ are Mandelstam variables and 
we used the following short-hands for the external spinors
$(\gamma^{\mu}P_{L,R})_{ee}\equiv\bar v(e^+)\gamma^{\mu}P_{L,R}u(e^-)$,
$(\gamma^{\mu}P_{L,R})_{\chi\chi}\equiv\bar u(\chi^+_i)\gamma^{\mu}P_{L,R} v(\chi^-_j)$.
The Sudakov corrections at leading order are written in full 
details in~\cite{NewReality} and can be expressed in a compact
form by splitting the amplitude as
$A_{ij}^{ab} = A_{ij}^{ab, H} + A_{ij}^{ab,W}$
where $a,b$ are the initial and final chiralities and $H$, $W$ stand
for the Higgsino and Wino components proportional 
to $Z_{2i}^{\pm}Z_{2j}^{\pm}$
and $Z_{1i}^{\pm}Z_{1j}^{\pm}$ respectively. Omitting for simplicity
the known renormalization group correction (of course, we included it in the 
analysis) the leading order Sudakov expansion of the amplitude reads
\ba
\label{Compact}
A_{ij}^{ab,H} &=& A_{ij}^{ab,H,Born}(1+\Delta A_{ij}^{ab,H}(q^2, \vartheta) + c_b^{Yuk}\log\frac{q^2}{M_W^2})
\nonumber\\
A_{ij}^{ab,W} &=& A_{ij}^{ab,W,Born}(1+\Delta A_{ij}^{ab,W}(q^2, \vartheta))
\ea
In the above expressions $\Delta A_{ij}^{ab}(q^2, \vartheta)$ is a logarithmic
correction fully described in~\cite{NewReality} depending on the c.m. energy and scattering angle $\vartheta$.
It contains the following parts: (i) a universal angular independent term proportional to the combination 
$2\log(q^2/M_W^2)-\log^2 (q^2/M_W^2)$, (ii) an additional term $\sim\log^2(q^2/M^2_W)$
in the Wino component only, (iii) an angular term proportional to  $\log(q^2/M_W^2)$
times a function of $\vartheta$.
The Yukawa contribution appearing in the Higgsino component depends on 
$\tan\beta$ through the coefficient
\be
\label{Yukawa}
c_b^{Yuk} = -\frac{3\alpha}{8\pi s_W^2 M_W^2}\left(\frac{m_t^2}{\sin^2\beta}\delta_{b,L}
+\frac{m_b^2}{\cos^2\beta}\delta_{b,R}\right)
\ee

Before we proceed, it seems opportune at this point to make our working strategy completely clear. As one
sees from Eqs.~(\ref{Compact}-\ref{Yukawa}) the logarithmic expansion contains as starting input the Born amplitudes of the
process. The latter ones depend on the \underline{bare} parameters $Z_{1i}^\pm$, $Z_{2j}^\pm$ that must be
expressed in terms of {\em physical} quantities defined in a chosen renormalization scheme. This would be
necessary in a conventional treatment at one loop, and we shall pursue the rigorous approach in a more
complete forthcoming paper. In this preliminary analysis, performed at next-to- leading (UV finite)
one-loop logarithmic order,
the problem disappears (it would affect the constant term of the expansion), and we can simply replace
the bare input with an identical {\em physical} one, whose meaningful definition will be given in the complete
one-loop treatment. 
The tree-level cross section depends on $\tan\beta$ but in the cases
we studied with $\tan\beta > 10$ and $M_2, \mu \gg M_Z$  this dependence
vanishes (actually, it will be generated in a sensible way in our description from the Yukawa vertex effects). 
Note that, still as a welcome consequence of our one-loop treatment, we can factorize
out all the QED IR divergent terms and consider them as a {\em known} quantity.

\begin{figure}[htb]
\centerline{\includegraphics*[width=8.5cm,angle=0]{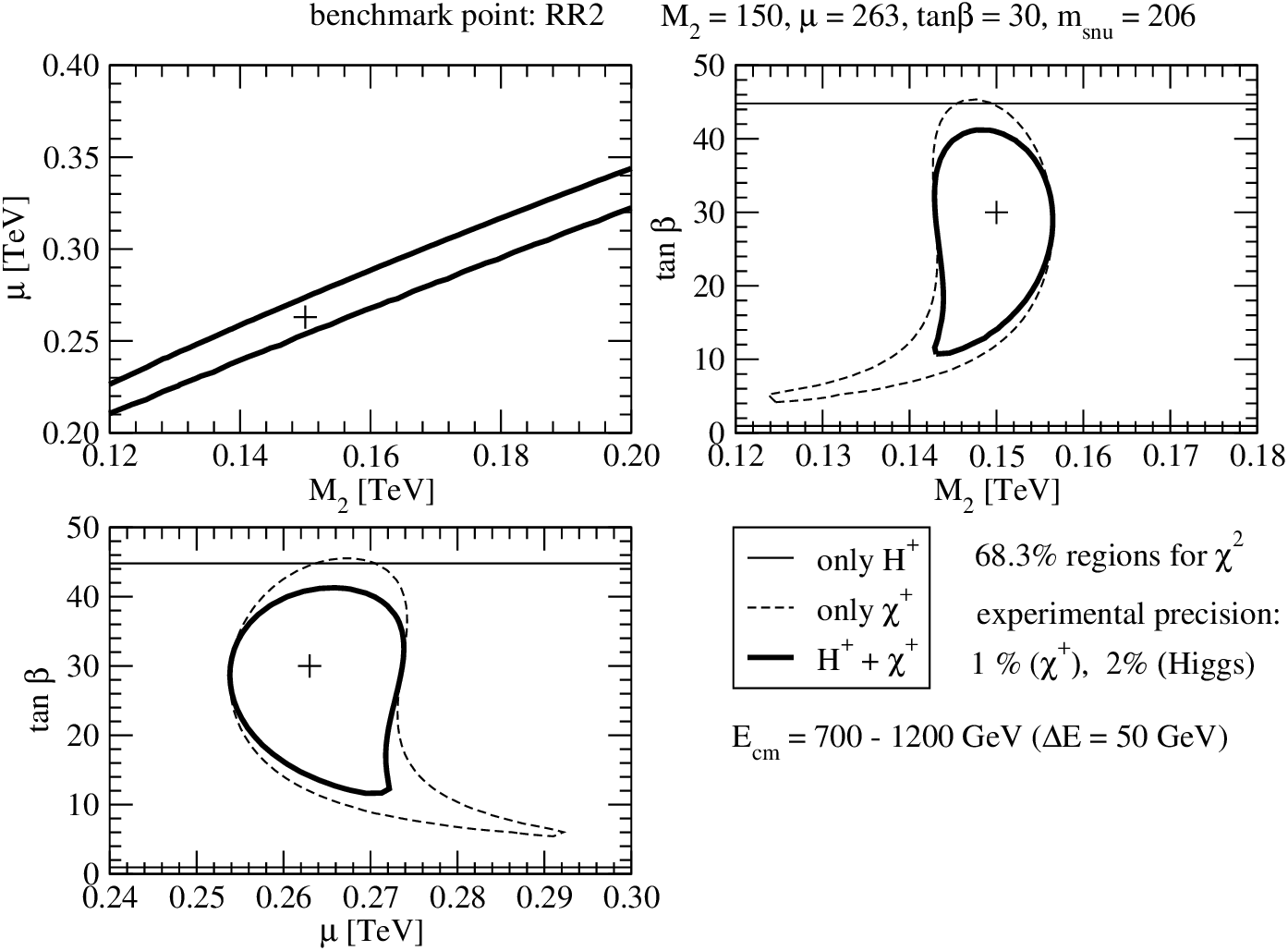}}
\caption{RR2 benchmark point. Bounds on the MSSM parameters $M_2$, $\mu$, $\tan\beta$.}
\label{fig:RR2a}
\end{figure}

\begin{figure}[htb]
\centerline{\includegraphics*[width=8.5cm,angle=0]{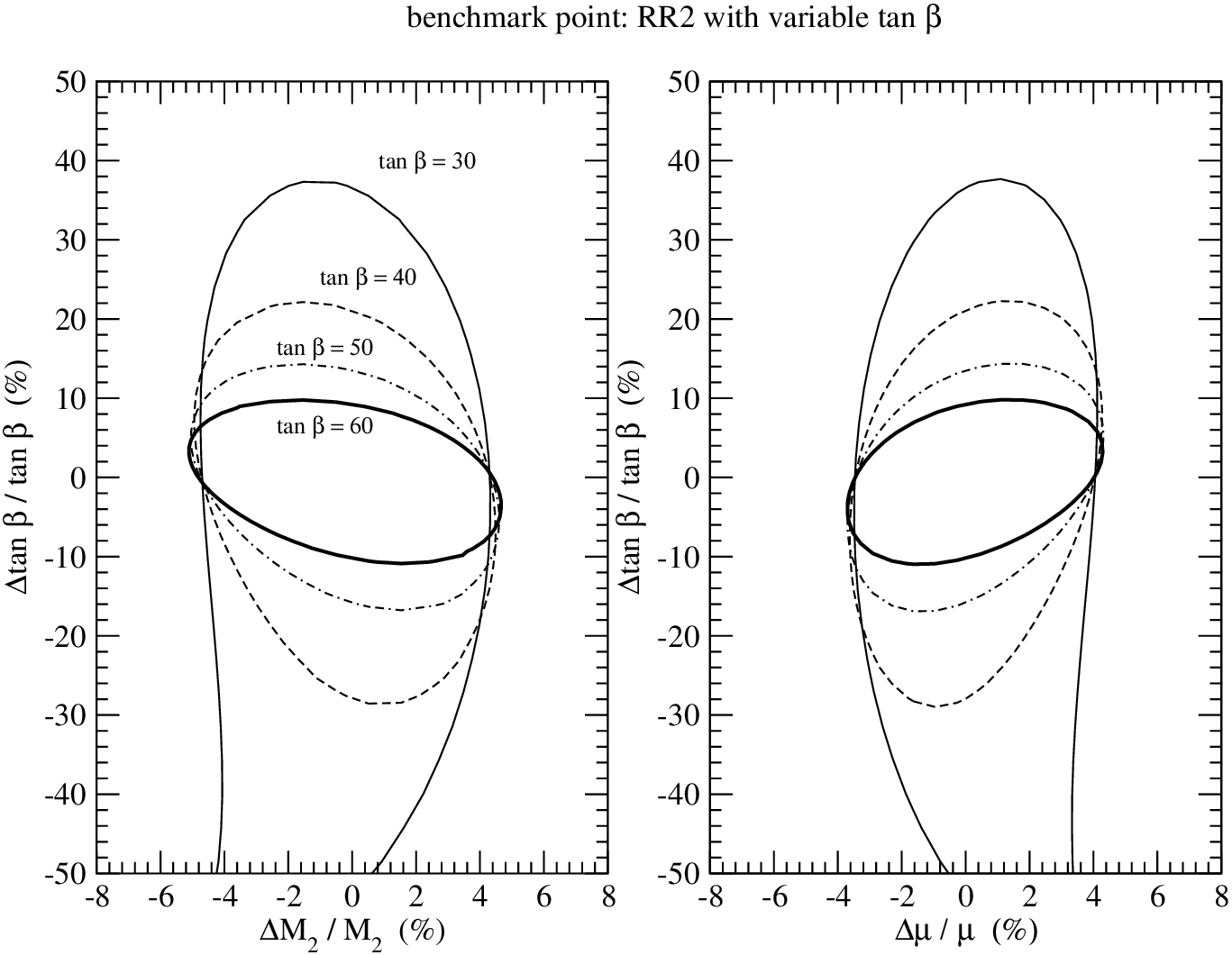}}
\caption{RR2 benchmark point. Relative bounds in the planes $(M_2, \tan\beta)$ and 
$(\mu, \tan\beta)$ as $\tan\beta$ is increased.}
\label{fig:RR2c}
\end{figure}

\begin{figure}[htb]
\centerline{\includegraphics*[width=8.5cm,angle=0]{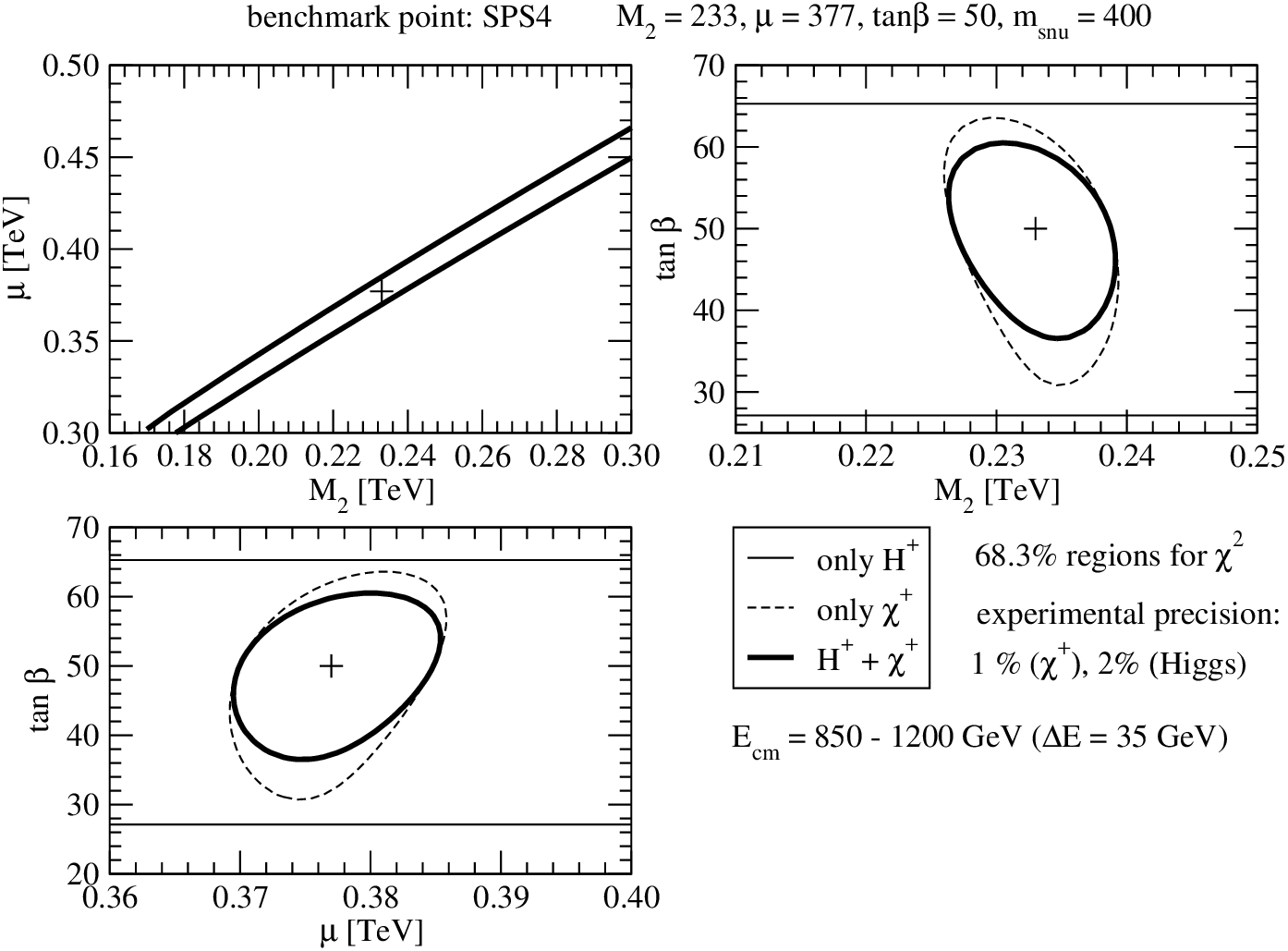}}
\caption{SPS4 benchmark point. Bounds on the MSSM parameters $M_2$, $\mu$, $\tan\beta$.}
\label{fig:SPS4a}
\end{figure}

\begin{figure}[htb]
\centerline{\includegraphics*[width=8.5cm,angle=0]{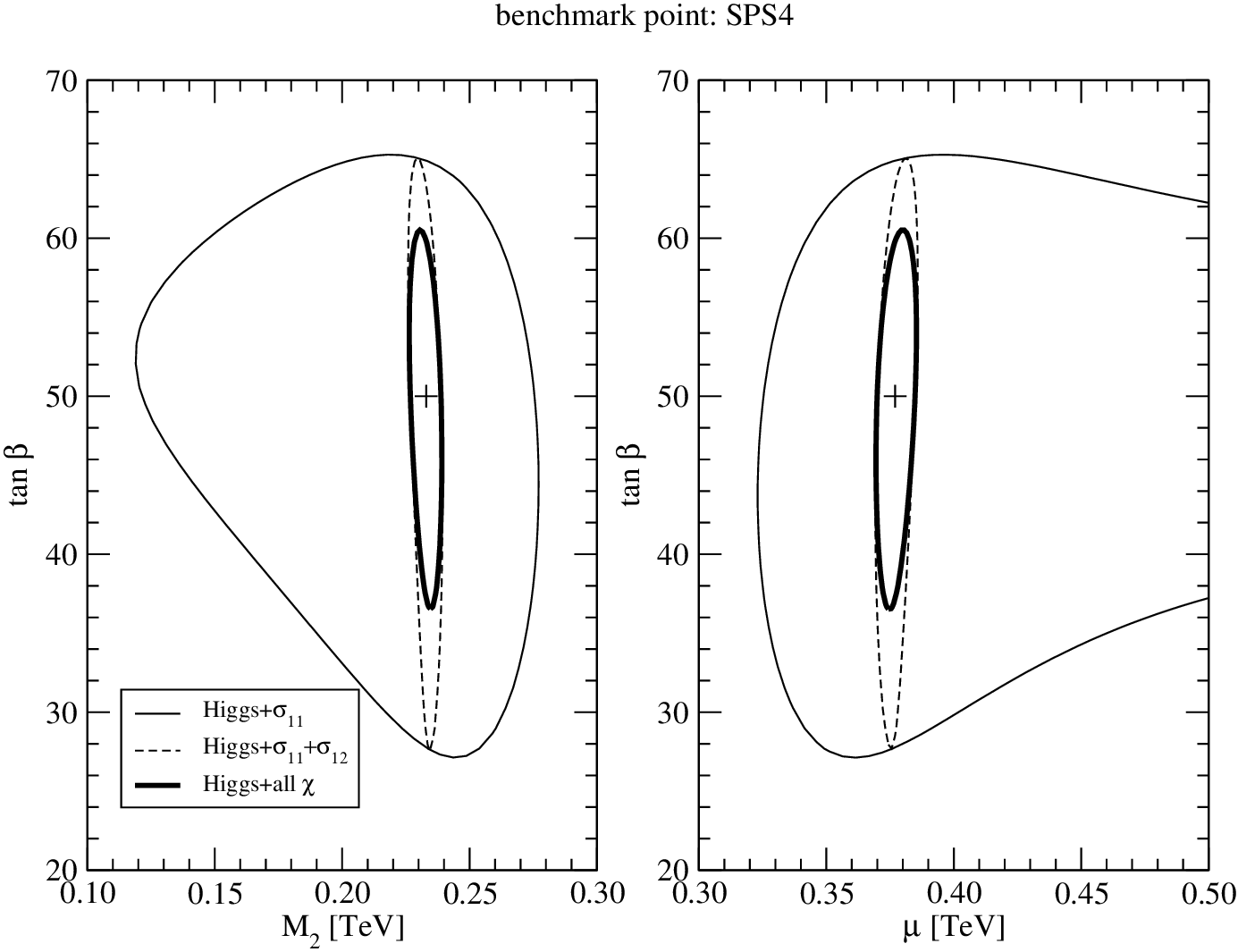}}
\caption{SPS4 benchmark point. Bounds on the MSSM parameters $M_2$, $\mu$, $\tan\beta$ 
with various combinations of cross sections.}
\label{fig:SPS4b}
\end{figure}

We performed a standard $\chi^2$ analysis of the Sudakov expansion of the 
cross sections for the above processes in order to determine bounds for the three
parameters $M_2$, $\mu$ and $\tan\beta$. With this aim, we assumed that a certain
number $N$ of cross sections measurements $\sigma(q_1^2), \dots, \sigma(q_N^2)$
are available and computed the quantities $q_n^2 \sigma(q_n^2) - q_1^2\sigma(q_1^2)$ 
where the unknown subleading constant in the Sudakov expansion cancels.
In Fig.~(\ref{fig:RR2a}), 
we show the results of the analysis at the TESLA
benchmark point RR2~\cite{RR2} where $M_2=150$ GeV, $\mu = 263$ GeV and $\tan\beta=30$. We 
took 10 points at equal distances for the four cross sections (pair production of 
$H^+H^-$ and three chargino channels $\chi_i^+\chi_j^-$
with $i,j = 11$, $22$ and $12+21$) at energies between 700 and 1200 GeV. The assumed experimental 
errors of this (optimistic) example are $1\%$ for charginos and $2\%$ for charged Higgs boson cross
section. The chargino masses
are consistent with the chosen energy range i.e. sufficiently smaller than the lowest considered energy
of 700 GeV. This is the only requirement in our approach while,
in a more conventional analysis, chargino masses would be input observables depending on a set of 
MSSM parameters larger than simply $M_2$, $\mu$ and $\tan\beta$.
The allowed region is a strip in the plane 
$(M_2, \mu)$ and comes from the parameter dependence of the mixing matrices in the chargino
cross sections. The bounds on $\tan\beta$ are conversely the combination of those coming from 
Higgses and charginos. They are mainly due to the Yukawa terms whose relative weight
in the Higgs case is bigger (roughly, a factor of two) than in the chargino cases (this motivates our choice of the experimental
uncertainties).
The relative errors on the three parameters in the planes $(M_2, \tan\beta)$ and 
$(\mu, \tan\beta)$ are shown in Fig.~(\ref{fig:RR2c}) where we also increased 
$\tan\beta$ to evidentiate the improvement due to the larger Yukawa contribution.
\noindent
\noindent
The same analysis can be repeated at the high $\tan\beta$ 
Snowmass benchmark point SPS4~\cite{SPS4} characterized by $M_2 = 233$ GeV, $\mu = 377$ GeV 
and $\tan\beta=50$. Now the heaviest chargino mass is around $400$ GeV and we performed the 
$\chi^2$ optimization starting at 850 GeV. The results are shown in Fig.~(\ref{fig:SPS4a}). 
\noindent
It may be interesting to analyze the different roles of the considered observables. 
With this aim we show in Fig.~(\ref{fig:SPS4b}) the bounds that can be 
derived with the Higgses and various subsets of the chargino 
cross sections.
\noindent

\section{Concluding Remarks}

 In the assumed logarithmic expansion, the slopes of the Higgs and of the chargino
cross sections depend only on three SUSY parameters i.e. $\tan\beta$, $M_2$, $\mu$. The parameter $\tan\beta$ appears in
both cross sections, with a relative weight that is more significant  in the Higgs process. To
improve the determination  that would be provided by the Higgs data alone, chargino data of
an experimental accuracy better than that achievable for Higgs data are requested, as we simulated in our 
 example. If this better
accuracy were not achieved, the relevant information on $\tan\beta$ in our proposed method would come almost
entirely from the Higgs pair production. For large values of the parameter, this would reach in any case interestingly small
error percentages, depending on the experimental accuracy and on the number of measured points in a way that it is
easy to calculate~\cite{ChargedUs} and that seems to provide a competitive determination of this fundamental
parameter~\cite{Ferrari}. On the contrary, the two remaining parameters ($M_2$, $\mu$) enter in the chargino cross
sections alone. Those measurements will not provide a separate determination of the two parameters, but will
force them to lie within an allowed strip in the $(M_2$, $\mu)$ plane. Thus, knowing the value of one of the parameters
will fix the other one, with a precision that we have shown in the (optimistic) accuracy assumptions that were
adopted (a more conservative strip can easily be drawn for different experimental assumptions). As we said in the
Introduction, our constraints on $\tan\beta$, $M_2$, $\mu$ could be obtained from a relatively simple analysis, in
particular from minimization programs that only contain these three quantities as unknown terms to be fitted. In
this sense, we believe to have shown in this (we repeat, preliminary) paper that from a combined investigation of
charged Higgs boson and chargino pair production at a future LC a strong consistency test of the considered
supersymmetric model would be, possibly and hopefully, derivable.

%
%

\end{multicols}

\begin{references}

\bibitem{LC} see {\em e.g.},
E.~Accomando {\it et.al.}, Phys. Rep. {\bf C299}, 299 (1998).

\bibitem{LEPSLC} The LEP Collaborations: ALEPH Coll, DELPHI Coll.,
L3 Coll., OPAL Coll., the LEP Electroweak Working Group, 
the SLD Heavy Flavor Working Group, hep-ex/0212036.

\bibitem{ChargedHollik}
J. Guasch, W. Hollik and A. Kraft, Nucl. Phys. {\bf B596}, 66 (2001).

\bibitem{ChargedUs} 
M. Beccaria,  F.M. Renard, S. Trimarchi, C. Verzegnassi,
Phys. Rev. {\bf D 68}, 035014 (2003).

\bibitem{NewReality}
M. Beccaria, M. Melles, F.M. Renard, S. Trimarchi, C. Verzegnassi,
Int. Jour. Mod. Phys. {\bf A18}, 5069 (2003).

\bibitem{Djouadi}
A. Datta, A. Djouadi, J-L. Kneur, 
Phys. Lett. {\bf B509}, 299 (2001).

\bibitem{Ferrari}
A. Ferrari, 
LC note LC-PHSM-2003-051, http:// www-flc.desy.de/ lcnotes/.

\bibitem{CharginoProduction} 
T. Blank, W. Hollik, hep-ph/0011092;
M.A. D\'\i az and D.A. Ross, JHEP {\bf 0106}, 001 (2001);
M.A. D\'\i az and D.A. Ross, hep-ph/0205257.

\bibitem{RR2} 
S.~Ambrosanio, G.~Blair, P.M.~Zerwas, 
in DESY ECFA 1998-99 LC Workshop, 
http://www.desy.de/ conferences/ ecfa-desy-lc98.html.

\bibitem{SPS4}
N. Ghodbane, H-U. Martyn, APS/DPF/DPB Summer Study on the 
Future of Particle Physics, Snowmass, Colorado, 30 Jun - 21 Jul 2001, 
hep-ph/0201233.

\end{references}
\end{document}